\documentclass{aastex61}

\PassOptionsToPackage{usenames}{color}

\usepackage{graphicx} 
\usepackage{comment}
\usepackage[normalem]{ulem}
\usepackage{hyperref}
\usepackage[usenames]{color}

\received{}
\revised{}
\accepted{}
\submitjournal{ApJ}

\begin{document}

\title{Solar Chromospheric Temperature Diagnostics: a joint ALMA-H$\alpha$ analysis }

\correspondingauthor{Kevin P. Reardon}
\email{kreardon@nso.edu}

\author{Momchil E. Molnar}
\altaffiliation{DKIST Ambassador}
\affiliation{National Solar Observatory, Boulder, CO, 80303}
\affiliation{Department of Astrophysics and Planetary Sciences,
University of Colorado, Boulder 
CO, 80303}

\author{Kevin P. Reardon}
\affiliation{National Solar Observatory, Boulder, CO, 80303}

\author{Yi Chai}

\affiliation{Center for Solar-Terrestrial Research, New Jersey Institute of Technology,Newark, NJ, 07102}

\author{Dale Gary}
\affiliation{Center for Solar-Terrestrial Research, New Jersey Institute of Technology,Newark, NJ, 07102}

\author{Han Uitenbroek}
\affiliation{National Solar Observatory, Boulder, CO, 80303}

\author{Gianna Cauzzi}
\affiliation{National Solar Observatory, Boulder, CO, 80303}
\affiliation{INAF-Osservatorio Astrofisico di Arcetri,
Florence, Italy, 50125}

\author{Steven R. Cranmer}
\affiliation{Department of Astrophysics and Planetary Sciences,
University of Colorado, Boulder 
CO, 80303}

\begin{abstract}

We present the first high-resolution, simultaneous observations of the solar chromosphere in the optical and millimeter wavelength ranges, obtained with ALMA and the IBIS instrument at the Dunn Solar Telescope. In this paper we concentrate on the 
comparison between the brightness temperature observed in ALMA Band 3 (3 mm; 100 GHz) and the core width of the H$\alpha$ 6563 {\AA} line, previously identified
as a possible diagnostic of the chromospheric temperature. We find that {in the area of plage, network and fibrils covered by our FOV} the two diagnostics are well correlated, with similar spatial structures observed in both.
The strength of the correlation is remarkable, given that the source function of the mm-radiation obeys local thermodynamic equilibrium, while the H$\alpha$ line {has a} source function {that} deviates significantly from the local Planck function. The observed range of ALMA brightness temperatures is sensibly smaller than the temperature range that was previously invoked to explain the observed width variations in H$\alpha$.

We employ analysis from forward modeling with the RH code to
argue that the strong correlation between H$\alpha$ width and ALMA
brightness temperature is caused by their shared dependence on the
population number $n_2$ of the first excited level of hydrogen. This population
number drives millimeter opacity through hydrogen ionization via the
Balmer continuum, and H$\alpha$ width through a curve-of-growth{-}like opacity
effect. Ultimately, the $n_2$ population is regulated by 
the enhancement or lack of downward Ly$\alpha$ flux, which coherently shifts the
formation height of both diagnostics {to regions with different temperature}, respectively.

\end{abstract}


\section{Introduction} \label{sec:intro}

Analysis of visible, UV, and infrared spectral lines has long provided a key method for extracting information about the solar chromosphere. The lines provide valuable information on velocity, magnetic fields, density stratification, and abundances, among other physical observables. However, due to the low-density conditions under which the chromospheric lines are formed, local thermodynamic equilibrium (LTE) typically does not hold. With an atmosphere whose structuring can be dominated by the local dynamics or the magnetic field topology, there is a possibility of steep gradients in density, temperature, or other parameters along the line-of-sight. Therefore, the interpretation of the information encoded in chromospheric line profiles is not straightforward.  Efforts at inversions of chromospheric spectral lines appear promising \citep[e.g.][]{2015A&A...577A...7S}, even though the robust extraction of atmospheric parameters from these profiles is still a subject of much research \citep{2019A&A...623A..74D, 2018A&A...617A..24M}.

Previous work has {suggested that the bisector width near the core of the Balmer-alpha transition line of hydrogen (H$\alpha$ from now on) may reveal information} about the temperature of the region of the chromosphere where {it forms} \citep{2009A&A...503..577C, 2012ApJ...749..136L}. 
The high temperature sensitivity of H$\alpha$ width was thought to primarily arise from the low atomic mass of hydrogen, resulting in a significant thermal Doppler broadening. 
{Furthermore, \citet{2009A&A...503..577C} found }
strong correlations between the H$\alpha$ line width and the width and core intensity of the \ion{Ca}{2} 8542 \AA~line{, which they} explained with a temperature dependent microturbulence, and a close coupling with local conditions (temperature), respectively.
However, the interpretation of the extreme widths of H$\alpha$ (reaching 1.2 \AA~and beyond) in terms of Doppler broadening implied temperatures of up to 50,000 K, which would cause the chromosphere to be fully ionized according to 1D models, and which are rarely found in 3D chromospheric models. 

The Atacama Large Millimeter Array \citep[ALMA;][]{2009IEEEP..97.1463W}, recently made available for solar observations, can provide an observable directly related to the electron temperature of the chromospheric plasma, supposedly freeing us from the non-LTE complications of other diagnostics \citep{2015ASPC..499..347P}. The continuum radiation at millimeter wavelengths ($\approx$ 0.3-10 mm)
originates from free-free emission in the chromosphere, and the two main opacity sources are the electron-ion free-free absorption and the
neutral-electron free-free absorption \citep[away from strong magnetic fields, see][for further discussion]{2016SSRv..200....1W}. These processes are coupled solely to the local properties of the plasma (electron temperature) and, therefore, result in a LTE source function. By using the Rayleigh-Jeans law, we can then interpret the emergent intensity in the millimeter wavelength domain as a local electron temperature. 
This was verified theoretically by \cite{2015A&A...575A..15L}, who showed that, in 3D MHD models, the ALMA brightness temperature indeed represents the electron temperature at the formation height of the millimeter radiation.

Early results of solar science with ALMA have been discussed by several authors. Among others, the visibility of chromospheric structures in full disk ALMA 1.21 mm data has been discussed by \citet{2018A&A...613A..17B}, while the presence and dynamics of chromospheric jets/spicules at the limb is reported by \citet{2018ApJ...863...96Y} 
and \citet{2018A&A...619L...6N}. Using high resolution IRIS observations of the Mg II h line obtained simultaneously, \citet{2017ApJ...845L..19B}
showed clearly the difference of temperatures derived from radiative diagnostics (Mg II line intensity) and the plasma temperature derived from ALMA. 

In this paper we concentrate on a joint analysis of the intensity measured in ALMA Band 3 and the H$\alpha$ line, to further our understanding of the chomospheric temperature structure. To this end, we employ some of the first simultaneous, high-resolution, high-cadence observations of the Sun in the millimetric range and in the optical and near-IR wavelengths.

\section{Observations}
\label{Ch:Observations}

\begin{figure*}
\includegraphics[width=\textwidth]{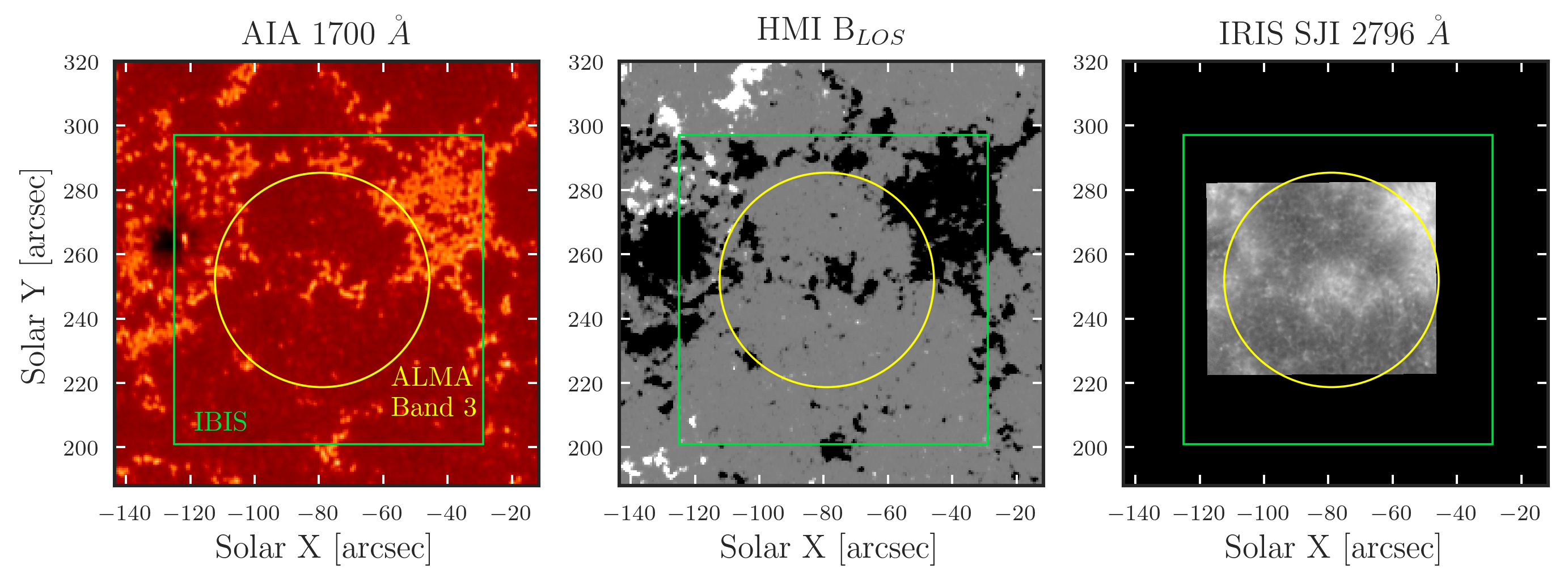}
\caption{Context images of the observed field of view from the following instruments at 17:25 UT: \textit{Left panel}: AIA 1700 {\AA} image; \textit{Central panel}: HMI LOS magnetogram, scaled (non-linearly) between -200 to 200 G.; \textit{Right panel}: IRIS SJI image at 2796 {\AA}, averaged over one minute. The field of view of IBIS is shown as the green square and the ALMA Band 3 field of view is shown as the yellow circle.}
\label{fig:context}
\end{figure*}

We obtained a coordinated set of observations between ALMA and the Dunn Solar Telescope \citep[DST,][]{1991AdSpR..11..139D} on April 23, 2017. At the DST, the Interferometric Bidimensional Spectrometer  \citep[IBIS, an imaging spectrograph][]{2006SoPh..236..415C,2008A&A...481..897R}, the Facility Infrared Spectrograph  \citep[FIRS, a scanning spectrograph,][]{2010MmSAI..81..763J}, and the Rapid Oscillations in the Solar Atmosphere instrument \citep[ROSA, a multichannel broad-band imager,][]{2010SoPh..261..363J} all observed the target region (in the following we concentrate only on the IBIS data). In addition, the IRIS \citep{2014SoPh..289.2733D} and Hinode \citep{2007SoPh..243....3K} satellites were co-pointing for these observations. Context images were available from SDO/AIA \citep{2012SoPh..275...17L} and magnetic field maps from SDO/HMI \citep{2012SoPh..275....3P}.

The observed target, shown in Figure~\ref{fig:context}, was an area of magnetic plage in the leading portion of NOAA active region 12651{, a stable region with low flaring activity present during the declining phase of the solar cycle}. Some quieter areas were present in the southern portion of the field of view (FOV).  The center of the target region was at E04, N11 at the time of the observations ($\mu = 0.96$ {or heliocentric angle of 16$^{\circ}$}). 
The primary leading spot of the active region was located about 60$\arcsec$ east of the target center, outside the field of view of all of the targeted observations. 

\subsection{DST/IBIS Observations}

 We observed the target region at the DST from 15:13 to 19:06 UT in conditions of good to excellent seeing. Although several different lines and spectral sampling combinations were obtained within the full observing interval, we focus here on a continuous series of IBIS observations that ran from 17:25 to 18:11 UT, and that included 180 interleaved scans of H$\alpha$ 6563, \ion{Ca}{2} 8542, and 
\ion{Na}{1} D1 5896 \AA. The lines were scanned with 29, 27, and 24 spectral sampling points, respectively, requiring between 3.4 and 4.0 seconds per line. With an additional overhead of 1.5 seconds to change prefilters, the total cadence for a scan of all three lines was 15.7 seconds. The spatial scale of the images from IBIS was $\sim 0.096\arcsec$/pixel. 

We applied linearity, dark correction, flatfielding, and fringe removal corrections to the IBIS data. In order to correct for optical and atmospheric image distortions, we employed a technique using the 
nearest-in-time HMI continuum intensity images in order to precisely map 
the IBIS spectral data onto a regular, fixed spatial grid with the bulk of the seeing distortions removed\footnote{using Rob Rutten's very capable software package
 available at \href{http://www.staff.science.uu.nl/~rutte101/Recipes_IDL.html}{his website}.}.

\bigskip

\begin{figure}

\includegraphics[width=\textwidth]{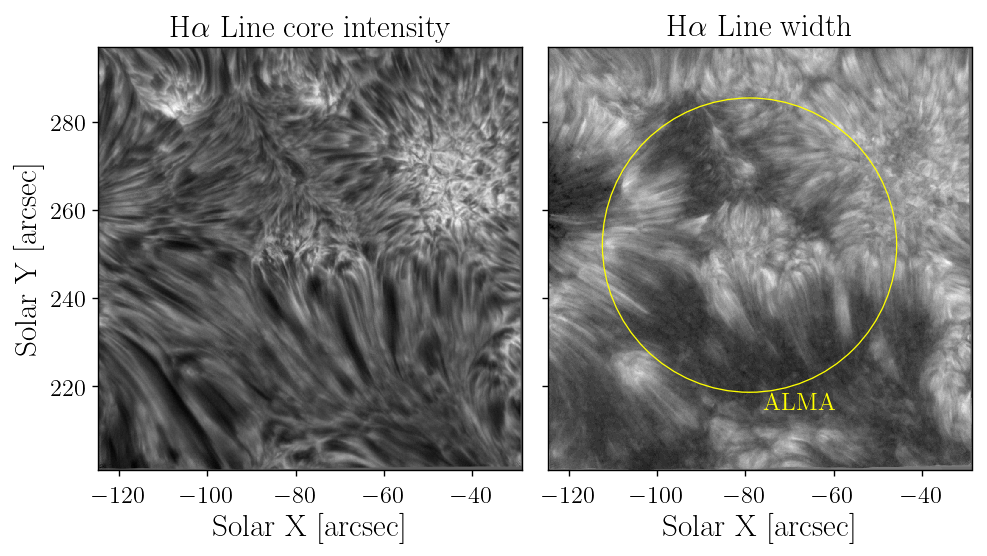}
\caption{\textit{Left panel}: H${\alpha}$ core intensity of the region observed by IBIS at 17:25 UT; \textit{Right panel:} H${\alpha}$ line core width,  scaled from 0.95 to 1.3~\AA {, with brighter pixels corresponding to relatively broader profiles.} The yellow circle shows the approximate ALMA field of view (see text for more details).}
\label{fig:ibis_parameter_maps}
\end{figure}

\subsection{ALMA Observations and Processing}
\label{sec:ALMA_processing}

The ALMA data were obtained with the array in configuration C40-3 with a maximum baseline of 460 m. However, due to antenna issues, during these observations the maximum baseline was 396 m. The data analyzed here were obtained in Band 3 (2.8-3.3 mm, 92-108 GHz) in the interval 17:19 to 18:53 UT.
The millimetric observations followed a sequence that dwelled for approximately 593 seconds on the target followed by a 145 second gap for observations of a phase calibrator. There were eight observing intervals all together, with the final interval being truncated to 440 seconds due to the end of the allocated observing block. After applying the standard radio interferometer data calibration procedures using CASA \citep{2012ASPC..461..849P}, we found that the ALMA images are heavily influenced by the phase disturbances due to water vapor variations in Earth's atmosphere, causing small-scale distortions on the images. To counteract this effect, we apply a self-calibration technique. For each 10 minutes observation interval, an average (clean) image is generated and then used to remove the random phase variations in the individual images within that observing interval. This allows us to minimize the image distortions in a self-consistent manner that does not introduce significant artifacts.

The field of view of ALMA band 3 images is only $\sim$ 60{\arcsec} diameter, which can be smaller than some of the large-scale structuring of solar features. To provide information on the background emission from the Sun, single-dish observations that scan the full disk were taken nearly simultaneously with the interferometer array \citep{2017SoPh..292...88W}, albeit with a significantly lower spatial resolution. Through the feathering process provided in CASA, we combine the two datasets covering the full range of spatial scales. In order to provide an absolute calibration for the measurements, the full-disk image was normalized such that the brightness temperature in the central region of the disk (diameter 190 arcsec) was equal to 7300 K, as determined by \citet{2017SoPh..292...88W}. This sets the background brightness temperature in our target region. This calibration compensates for any absorption variations in the terrestrial atmosphere and provides an absolute temperature accurate to 2-5\% \citep{2017SoPh..292...88W}.

Figure \ref{fig:ALMA_vs_IBIS} (top row) shows examples from the final results after the calibration and feathering procedures. The effective field of view of these Band 3 observations was approximately 60{\arcsec} in diameter; the restoring beam size of the ALMA image, as calculated in CLEAN, is 1.75{\arcsec} x 1.91{\arcsec} in the x-y directions. The shape of synthesized beam depends on the antenna configuration and the angle between the array and the target, which means it is not circularly symmetric. While the beam size will change size and shape slightly over time, our observing interval was short enough that there were no significant changes to the beam shape.

\section{Diagnostics Comparison}
\label{Ch:Diagnostic_Comparison}

\subsection{IBIS Line Widths}
\label{sec:IBIS_line_widths}
We concentrated on the diagnostics afforded by the H$\alpha$ line, calculating several parameters from the spectral profiles recorded at each pixel and for each time step. The intensity and wavelength of the line-minimum position were determined through fitting of a second-order polynomial to the line core. {We calculated the width of the line core following the technique described in \cite{2009A&A...503..577C} -- in short, we measure the separation of the line profile wings at half of the line depth, defined as the difference between the line minimum and the intensity at $\pm$ 1.0 \AA~from the core position.} This determines the line width in the central core of the line profile, which is essentially the portion formed at chromospheric heights. The instrumental profile has a negligible effect on the measured width, and the errors in the line width were determined to be about 0.005~\AA.

Intensity and line width maps for the data acquired at 17:25~UT are shown in Figure~\ref{fig:ibis_parameter_maps}.
While the H$\alpha$ core intensity shows the familiar ``forest'' of chromospheric fibrils originating from magnetic features and covering much of the internetwork regions (compare Figure~\ref{fig:context}), the line width map shows a significantly different scene. The plage and magnetic network are clearly highlighted as region of large line widths, together with the base and partial length of a selected set of fibrils, as discussed in \cite{2009A&A...503..577C}.  There is fine spatial structuring in the line width, down to the resolution limit of the data.

\subsection{ALMA Band 3 Intensity}
\label{sec:ALMA_Band_3_intensity}

The ALMA Band 3 intensity maps are shown in the top row of Figure~\ref{fig:ALMA_vs_IBIS}. 
As stated above, the resolution of the ALMA images is $\approx$ 2.0$\arcsec$. 

The top left panel in Figure~\ref{fig:ALMA_vs_IBIS} displays a single 2 second ALMA integration, whereas the top right panel is a time average over the 10-minute time block observed between 17:31~UT and 17:41~UT. 

From the similarity of the instantaneous and time-averaged maps, it appears that at this spatial resolution the structures remain fairly stable over several minutes, although ALMA movies show subtle evolution; a complete temporal analysis is left for a future work. 

The 3 mm intensity maps are clearly brighter (hotter) in network and plage regions, with $T_B$ reaching up to $\sim$ 12,000~K in the latter, whereas in the internetwork temperatures are as low as $\sim$ 6,500 K. The overall average $T_B$ for our field is $\sim$ 8600 K, reflecting the presence of plage in the FOV; the average in the quieter portions of the FOV is $\sim$ 7500 K. This is consistent with the quiet Sun calibration described in Sec. \ref{sec:ALMA_processing}. 

The most interesting property of the ALMA $T_B$ maps, however, might be the presence of several bright ALMA features that appear fibrilar in nature, reminiscent of those observed in the H$\alpha$ line width map of Figure \ref{fig:ibis_parameter_maps}. They are particularly prominent in the bottom part of the FOV, where some of the strongest H$\alpha$-width features are seen.  To the best of our knowledge, this is the first time that fibrilar structures are clearly identified in on-disk ALMA images. Contrary to the prediction of %\citet{2017IAUS..327....1R}, 
\citet{2017A&A...598A..89R}, however, we identify them as bright (hence hot) features, rather than dark, strong fibrils coincident with H$\alpha$ line-core intensity features.

\subsection{Comparison between the ALMA Band 3 intensity and H$\alpha$ line width}

\begin{figure}

          \includegraphics[width=\textwidth]{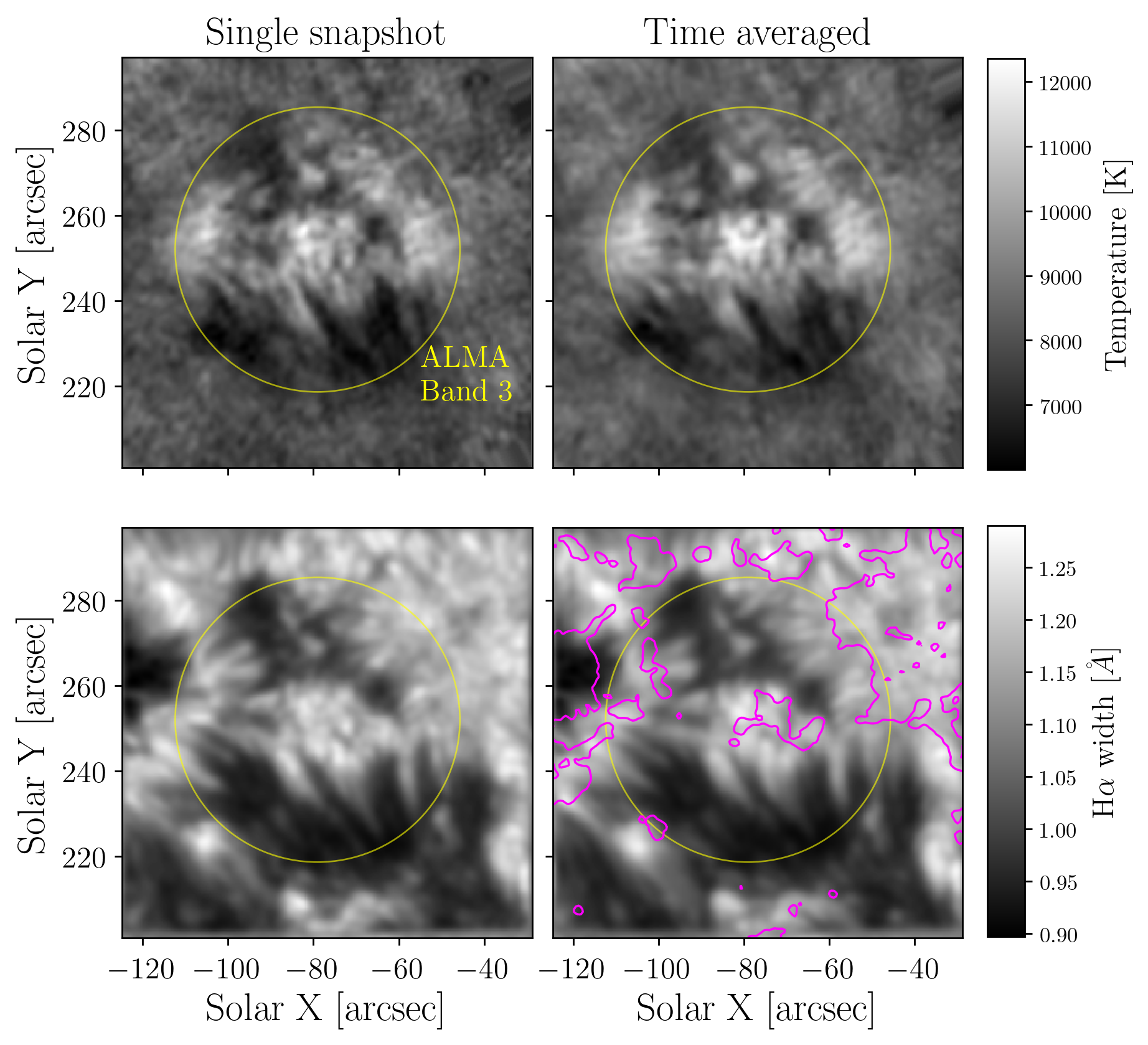}
\caption{Maps of ALMA intensity (top row) and IBIS H$\alpha$ line width (bottom row). The left column shows the two parameters at single time step of the observations, while the right column shows the same two parameters averaged over the ten minutes of a continuous ALMA observation block. The IBIS data have been smoothed to match the resolution of the ALMA data, by using a elliptical Gaussian kernel with FWHM in the x-y directions of 1.95~\arcsec $\times$ 2.03~\arcsec. The yellow circle with a diameter of 66~\arcsec, is slightly larger than the usable field-of-view of ALMA. The {purple} contours overlaid on the bottom right panel show areas with magnetic field strength above 500 Gauss measured by HMI.
}
\label{fig:ALMA_vs_IBIS}

\end{figure}

The bottom panels of Figure~\ref{fig:ALMA_vs_IBIS} show the maps of H$\alpha$ width {acquired cotemporally} to the ALMA maps: a single snapshot in the left column; and the average over the period 17:31 $-$ 17:41 UT in the right column. Contours of the HMI longitudinal magnetic field magnitude at values of greater than 500 Gauss have been drawn in yellow on the right panel to facilitate comparison with Figure~\ref{fig:context}. 
In order to provide a more relevant comparison, we have degraded the spatial resolution of the IBIS data to best match the appearance of the ALMA data. We find that the minimum difference between the blurred H$\alpha$ width and ALMA images occurs for a circular Gaussian with a FWHM 2.0\arcsec. This is about 8\% greater than the mean FWHM of the theoretical ALMA beam, likely due to residual atmospheric smearing at the millimeter wavelengths. 
 Since the ALMA PSF is known to be non-circular due to the orientation of the interferometer array with respect to the object, we apply the ratio between the x and y axes of the ALMA beam (as calculated for our observations by CLEAN) and convolve the IBIS images with an elliptical Gaussian kernel having FWHM in the x-y directions of 1.95\arcsec $\times$ 2.03\arcsec, respectively.
In fact, the match is so satisfactory that we were able to align the H$\alpha$ width maps and ALMA images to easily achieve sub-arcsecond accuracy. 
The alignment between the two diagnostics was done on the time averaged images and, given the spatial resolution, did not appear to produce any artificial superposition of localized features. Because the temporal sampling was higher for ALMA compared to the multi-line IBIS scans (2 vs. 15 seconds), we temporally binned the ALMA data to match the IBIS temporal sampling by combining the eight closest-in-time ALMA brightness maps.

The similarity between ALMA intensity and H$\alpha$ width is striking, with regions of similar size, shape and contrast. High/low brightness temperature in ALMA Band 3 correspond, almost one-to-one, to broad/narrow H$\alpha$ profiles. The HMI contours make clear how the ALMA fibrilar structures represent the lower portion of heated features originating from regions with high magnetic flux, like those seen in H$\alpha$.

\begin{figure}
\begin{center}
\includegraphics[width=\textwidth]{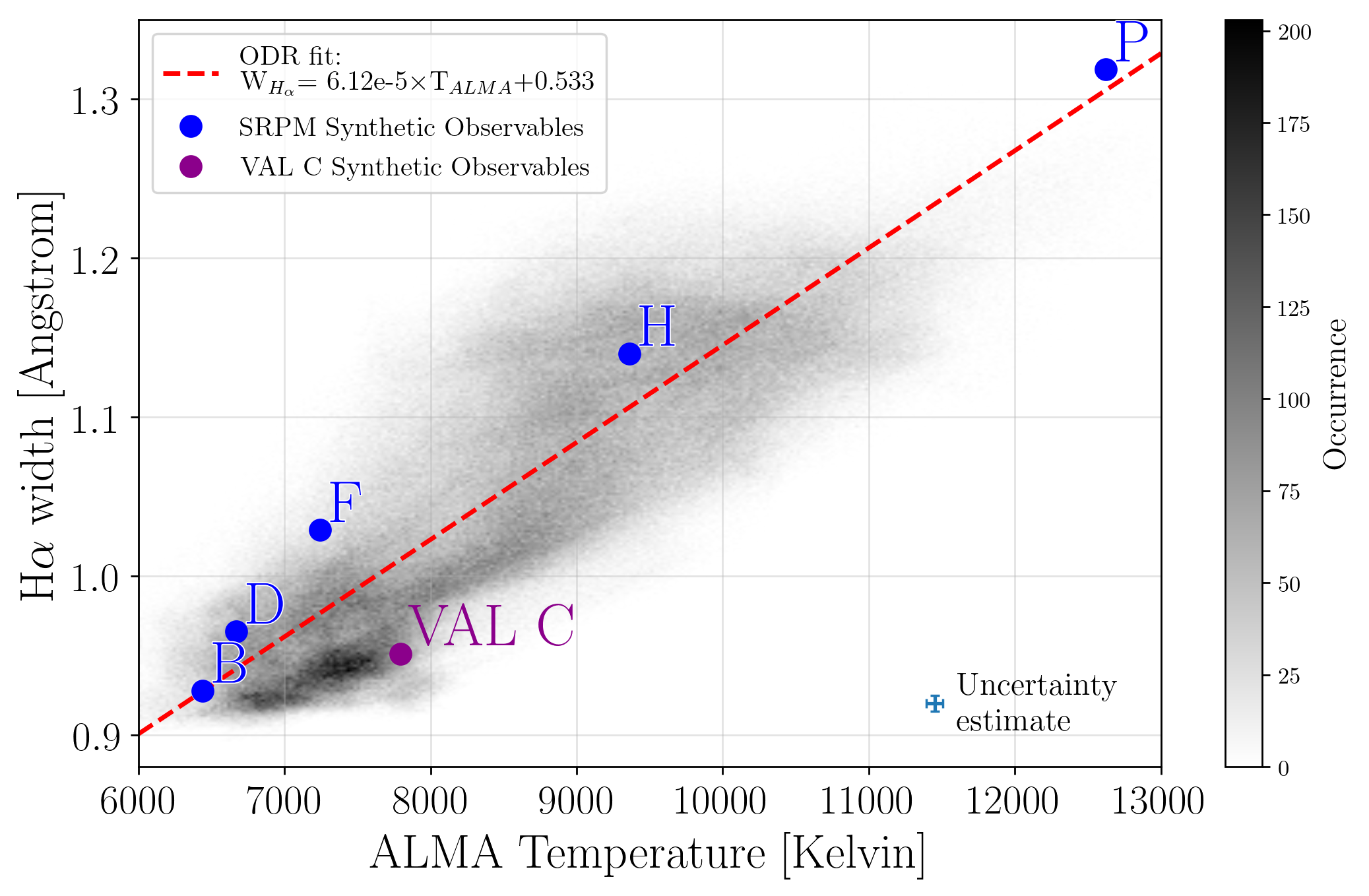}
\caption{2D histogram of H${\alpha}$ core width versus ALMA brightness temperature, clearly showing the correlation between the two 
quantities. Overlaid as blue dots are the result from the RH spectral synthesis in the {SRPM atmospheres and in violet in the VAL C model}.  The uncertainty estimates in both directions are presented with the marker in the bottom right of the plot. The spread in 
the figure is physical, as it is significantly larger than the uncertainties in our measurements. Orthogonal distance regression (ODR) was used to fit the observation with a straight line, resulting in the red dashed line.}

\label{fig:IBIS-ALMA_scatter}
\end{center}
\end{figure}

To quantify the comparison, a 2D histogram between the temporally coincident maps of ALMA Band 3 brightness temperature and H$\alpha$ width is shown in Figure~\ref{fig:IBIS-ALMA_scatter}. The {histogram derived from} the {overall} time-average of {these} two diagnostics shows an essentially identical distribution. The correlation is very strong across the full range in temperature {present in our FOV}, with a Pearson correlation coefficient of 0.84. Because of the scatter present in both measurements, we perform an orthogonal distance regression \citep{1990ApJ...364..104I} to find the best linear fit between the two parameters. The best fit equation is $0.533 + 6.12\times10^{-5}\;T$, which is plotted in Figure~\ref{fig:IBIS-ALMA_scatter}.

However, the range of T$_B$ measured with ALMA is not large enough to explain the change of the observed H$\alpha$ line width from 0.95 to 1.2 \AA\ as solely due to thermal Doppler broadening, as hypothesized by \citet{2009A&A...503..577C}. This is because the H$\alpha$ line width already has a significant intrinsic width of around 0.95 \AA\, and this dominates the quadratic sum of broadening components for the observed range of {ALMA} temperatures. In fact, an increase from 6,000 to 12,000 K (Figure~\ref{fig:ALMA_vs_IBIS} top row) would only be expected to increase the line width by 0.025 \AA, an order of magnitude smaller than the observed broadening.

Finally, we note the presence of a cluster of points between {6,500-7,500 K for which the H$\alpha$ widths fall 0.03 to 0.05 \AA\ lower than the fitted correlation.}
These points correspond to quieter regions in the {bottom half of the} ALMA FOV, {and located farther away from the magnetic concentrations}.

\section{Synthetic Spectral Diagnostics with RH}
\label{Ch:Radiative_transfer}
\begin{figure}[ht]
\includegraphics[width=\textwidth]{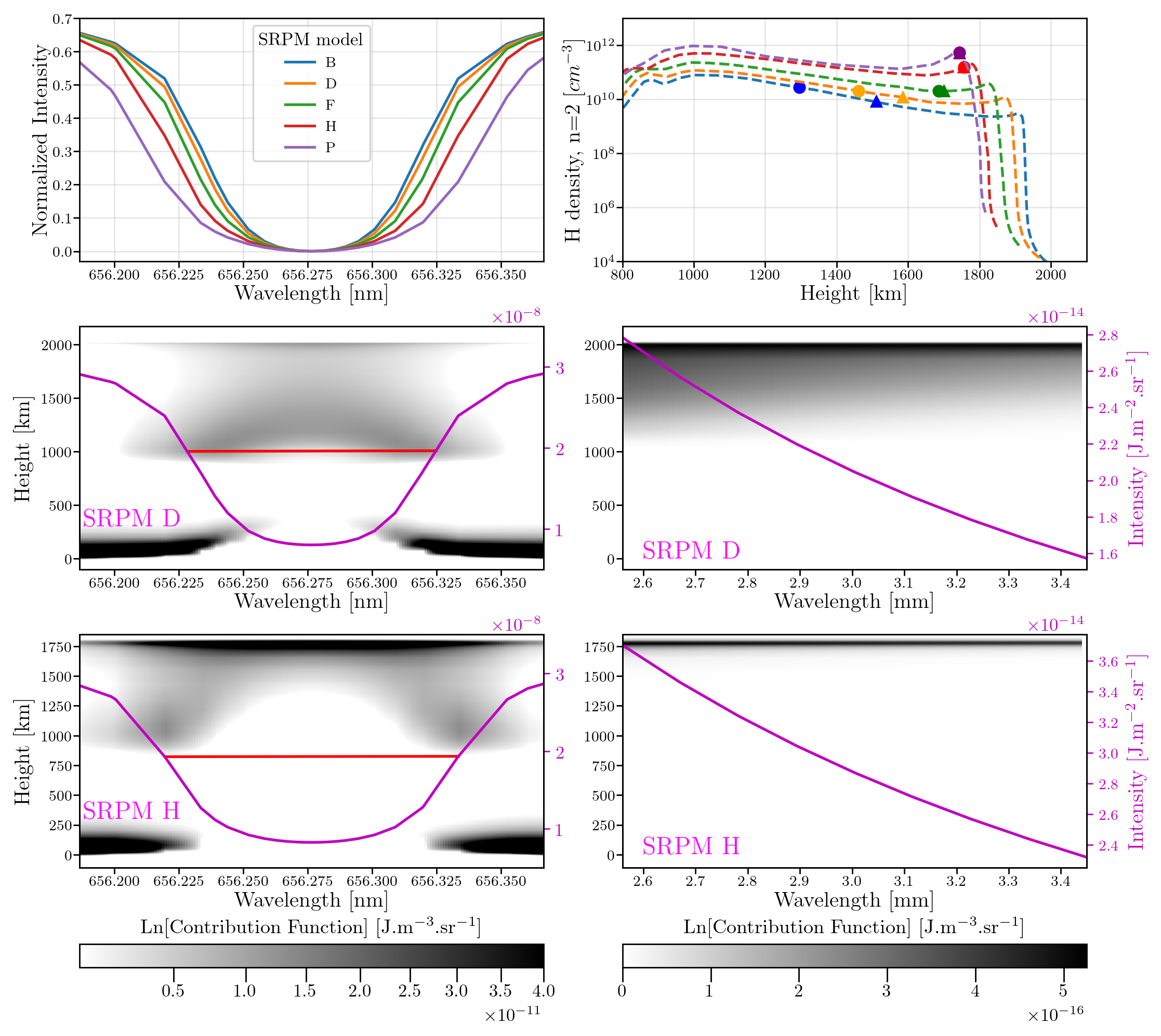}
\caption{Results from the RH spectral synthesis. \textit{Upper left}: H$\alpha$ profiles for the different {SRPM} atmospheres. \textit{Upper right}: Number density of hydrogen atoms in the $n=2$ quantum state. The symbols show the height at which $\tau=1$ for the 3~mm radiation (triangles) and the H$\alpha$ line core intensity (circles), in the corresponding atmosphere. {Note that the triangles and circles indicating the $\tau=1$ surfaces coincide for the hotter models.}  \textit{Middle left}: intensity contribution function for the H${\alpha}$ line for the {SRPM} D model overlaid with the emergent line profile (\textcolor{violet}{in violet}). Note the different intensity scaling with respect to the upper panel. {The position of the line width measurement described in Section~\ref{sec:IBIS_line_widths} is illustrated with the red line.}  \textit{Middle right}: contribution function for the emergent intensity for
ALMA Band 3 wavelengths for the {SRPM} D model, overlaid with the emergent intensity profile (\textcolor{violet}{in violet}). \textit{Bottom left, right}: as the middle panels, for the {SRPM} H model. 
}

\label{fig:RH_results}
\end{figure}

To investigate the mechanisms behind the correlation of Figure \ref{fig:IBIS-ALMA_scatter}, we utilized the RH code \citep{2001ApJ...557..389U} to synthesize observables for different solar atmospheric models. We chose the 1D {Solar Radiation Physical Modeling (SRPM from now on)} models  \citep{2011JGRD..11620108F} as input for our work. We used models ranging from the Quiet Sun internetwork (model B)
to bright facular region (model P), representative of the different
solar structures in our field of view. {We also included the heritage quiet Sun VAL C model for comparison \citep{1981ApJS...45..635V}.} 
The RH calculations were done using a 4-level (including continuum) hydrogen atom {with the Ly-alpha, Ly-beta and Balmer-alpha transitions treated under partial redistribution (PRD)}, while the rest of the atomic species were treated in LTE.  
The widths of the chromospheric core of the synthetic H${\alpha}$ profiles were measured using the same method used for the observational data. The emergent millimetric radiation was calculated from the synthetic intensity under the Rayleigh-Jeans approximation for a wavelength range  of 2.6-3.4 mm (100 GHz), which covers the observed wavelength interval.

The {SRPM} models have shortcomings as they are 1D, hydro-static, semi-empirical models optimized to reproduce the temporally averaged solar spectrum as observed at a few arcsecond resolution. Yet, given that the data of Figure~\ref{fig:IBIS-ALMA_scatter} has been smoothed to a similar resolution, we
use them as a first attempt to provide some physical insight. It is also worth noticing that modern, ab-initio 3D--MHD atmospheric models such as BIFROST~\citep{2011A&A...531A.154G, 2016A&A...585A...4C}
still lack some significant physical processes in the range of heights of relevance for our work; in particular, they do not reproduce the correct width of chromospheric lines \citep{2009ApJ...694L.128L}, which is obviously crucial for our analysis. In addition, it would be more correct to perform the radiative transfer calculations for H$\alpha$ using full, 3D radiative transfer computations~\citep{2009ASPC..415...87L}, but we leave that to a future work due to complexity and computational requirements.

\begin{comment}
\begin{table}[h!]
\begin{center}
\begin{tabular}{ c c  } 
 \hline
 FAL model label & Atmospheric description \\ \hline
 A & Dark Quiet-Sun inter-network \\ 
 B & Quiet-Sun inter-network\\ 
 C & Quiet-Sun network element (see \cite{1993ApJ...406..319F})\\
 D & Quiet-Sun network element \\
 F & Enhanced network \\
 H & Plage \\
 P & Facular region \\
 \hline
 
\end{tabular}
\caption{SRPM atmospheric models used for the RH synthesis as described in \cite{2011JGRD..11620108F} and model C from \cite{1993ApJ...406..319F}.}
\label{tab:FAL11_models}
\end{center}
\end{table}
\end{comment}

The results from the RH line synthesis are presented in Figure~\ref{fig:IBIS-ALMA_scatter} as blue circles, and coincide well with the observational data
{for models B to P}, 
 which are the dominant features in our observations
{(note that hotter models, e.g. model Q, do not follow the trend shown in Figure \ref{fig:IBIS-ALMA_scatter}, hinting at a perhaps different behavior for other solar features like active regions).} 
{The VAL C result falls below the observed correlation, with too small of a line width, likely due to some of the simplified physical assumptions in that older model (e.g lack of ambipolar diffusion and a different treatment of Ly$\alpha$).}
The range of ALMA temperatures {reproduced by the models} is not surprising, since the semi-empirical models were tailored to reproduce observed millimeter continuum brightness, among other diagnostics. The observed range of temperatures and correspondence to the appropriate chromospheric structures in the field confirms the proper calibration of the ALMA brightness temperature and consistency with previous work. However, the SRPM models were not constructed using H$\alpha$ line widths as a parameter, which makes the close correspondence especially pleasing.

Using the RH results, we can investigate the formation of H$\alpha$ and the millimetric radiation in more detail. Figure~\ref{fig:RH_results} shows the run of relevant parameters for different atmospheric models.
In the top left panel, we see how the H$\alpha$ line saturates even in the colder models, and grows broader as the chromospheric temperature rises. Interestingly, this appears directly correlated with sensibly higher populations of the hydrogen $n=2$ level in the hotter models, as plotted in the top right panel. Indeed, the number density of H atoms in the $n=2$ level increases two orders of magnitude from model B to model P, suggesting that the broadening of H$\alpha$ could be primarily due to an opacity effect (akin to a curve-of-growth plot).

The middle and bottom panels of Figure~\ref{fig:RH_results} show the intensity contribution functions (CFs) for both diagnostics, for the representative models D and H.  The emergent intensities over both the H$\alpha$ profile and the ALMA band 3 wavelengths are plotted in violet {(in the left panels we also show the width of the resulting H$\alpha$ line, as computed with our method marked with the red line)}. These panels show that while both diagnostics form over the same general expanse of the chromosphere, their range of formation becomes narrower and more coincident as we move to hotter models. This is represented in the top right panel, where the symbols indicate the heights at which the optical depth $\tau=1$ for the 3~mm radiation (triangles) and for H$\alpha$ line core (circles). 

For model H, a large fraction of the H$\alpha$ core intensity, and essentially all of the millimetric intensity originate from a narrow region at the interface  of the chromosphere and the transition zone. For the colder model atmospheres the height separation between the two diagnostics becomes significant. However, some correlation between the diagnostics is still to be expected because both temperature and the hydrogen $n=2$ level populations (see below) vary very slowly in the relevant range of heights.

We hypothesize that the underlying physical mechanism for the correlation shown in Figure~\ref{fig:IBIS-ALMA_scatter} could arise from the common sensitivities of both diagnostics
to the population numbers in the first excited state of hydrogen, $n_2$, in particular as related to excitation in Ly$\alpha$. Most excitations in Ly$\alpha$ occur
via radiative transitions, as not many electrons have sufficient
energy to collisionally excite the line at chromospheric temperatures.
Opacity in the line core is very high and the radiation field is completely
thermalized. However, there is significant excitation in the line
wings by radiation coming from the hot transition region above
\citep{2003ASPC..286..403U,2012ApJ...749..136L}. Being this far in the UV, the additional radiative excitation through downward
flux of Ly$\alpha$ wing photons is very sensitive to the {column mass at which the transition region occurs}. 
In Figure~\ref{fig:RH_results} (upper right) the effect of the downward radiation in
the Ly$\alpha$ wings is visible as the bump on the right side of the
$n_2$ level population plot, on top of the general rise from one
model to the next that is caused by the increase in density scale height
associated with hotter models. 

On the one hand, an enhanced value of $n_2$ leads to enhanced ionization of hydrogen,
as the predominant mechanism for hydrogen ionization in the chromosphere
is via the Balmer continuum, which decouples from local conditions
already much deeper in the atmosphere \citep{2002ApJ...572..626C}. Since the radiation field
in the Balmer continuum is optically thin, any fractional increase
in the $n = 2$ populations will raise ionization levels proportionally.
The increase in electrons coming from this increase in $n = 2$ population raises the
opacity at 3 mm, moving the formation height of radiation at these
wavelengths up in the atmosphere to higher temperatures; this
results in raising the millimetric brightness temperature, once 
hydrogen ionization reaches a few percent.
 
On the other hand, an increase in $n_2$ raises the opacity in H$\alpha$,
also increasing the formation height of the core in particular.
The line source function of H$\alpha$ is almost flat with
height through the chromosphere, resulting in the characteristic
flat bottom of the line profile in the core. Changing wavelength
from the core outward in the line profile the intensity follows
the flat source function inward
\citep{2012ApJ...749..136L, 2012A&A...540A..86R}, until it suddenly becomes sensitive to the
photospheric temperature rise with depth, resulting in the steep
wings of the line profile. The further the line core formation moves up in {the
atmosphere}, the further in wavelength we have to move out of the core
to see the wings, explaining the dependency of the H$\alpha$
width as an opacity effect, rather than a direct effect from
thermal broadening.
 
Thus, we suggest that both the 3 mm brightness temperature and the
H$\alpha$ line width depend coherently on the {$n_2$ level populations of hydrogen in the chromosphere. With higher values of $n_2$ the formation height of the 3 mm radiation increases
through increased hydrogen ionization in the Balmer continuum as
explained above. At the same time, an increase in $n_2$ leads to opacity
broadening of H$\alpha$ through an upward shift of the core
formation height (as noted above, for the hotter SPRM Q the trend breaks down as  H$\alpha$ forms at the base of the TR where the higher source function leads to an increased line core intensity), and the effect of an almost flat source function of the line 
through the chromosphere. The $n_2$ population of hydrogen in the chromosphere is determined by the downward Ly$\alpha$ wing flux \citep{2002ApJ...572..626C}, and, in turn, the Ly$\alpha$ flux is determined by the column mass at which the TR occurs \citep{1992str..book.....M}. Indeed, in our synthesis we found increasing Ly$\alpha$ flux for the hotter models (B to P), whose TR location occurs at increasing column mass. This is in agreement with our conjecture about the correlation between H$\alpha$ width and ALMA brightness temperature.}

\section{Conclusions} 
\label{Ch:Conclusion}

We have presented the first observations combining  high-spatial-resolution spectral imaging in the traditional chromospheric indicator of H$\alpha$  combined with simultaneous brightness temperature maps  at millimeter wavelengths obtained with ALMA. The common, $\sim$60{\arcsec} diameter FOV of the two instruments contained plage, network, and some (magnetically) quieter areas. 

The ALMA 3-mm images display a structured pattern of bright (hotter) and dark (cooler) features, with spatial sizes down to the 
spatial resolution of $\sim$2.0 $\arcsec$. The corresponding brightness temperature spans a range between 6,500 and 12,000 K. An interesting
property of the ALMA $T_B$ maps is the presence of bright ALMA features that appear fibrilar in nature, particularly prominent in the 
bottom part of the FOV. The ALMA images bear a striking similarity with the maps of the H$\alpha$ line core width, with features of similar 
size, shape and contrast (Figure~\ref{fig:ALMA_vs_IBIS}). This is contrary to the predictions presented in a recent paper by  
\citet{2017A&A...598A..89R},
that hypothesized the ubiquitous presence of long, opaque ALMA fibrils, with a good dark-dark correspondence with the H$\alpha$ core intensity; as shown also in 
\citet{2009A&A...503..577C}, the H$\alpha$ core intensity and width are poorly correlated quantities. We however defer a more detailed comparison of such features to a further work.

Our most important result is the strong quantitative correlation between the intensity of ALMA and the width of the H$\alpha$ line core {in the range of observed temperatures}, as plotted in Figure~\ref{fig:IBIS-ALMA_scatter}. Using forward synthesis with RH, we showed that the correlation is well reproduced with 1D semi-empirical models of typical solar structures, which further indicates that the H$\alpha$ intensity and the millimeter radiation are formed in a similar span of the middle-upper chromosphere.
{We note that the synthesis from the VAL C model falls below the observed correlation and away from the other modeling results, most likely due to the inclusion in the SRPM models of ambipolar diffusion and a different, more detailed treatment of the Lyman-$\alpha$ line profile.}

The main factor driving the correlation appears to be that 
the opacity sources for both spectral diagnostics is determined through the $n=2$ hydrogen population. The mm-wavelength opacity depends on the electron number density, which is related to the $n=2$ population of hydrogen as the statistically dominant source of free electrons in the upper chromosphere. At the same time, the line-broadening of H$\alpha$ is determined by the column mass of $n=2$ hydrogen atoms in a manner similar to a curve-of-growth effect. This effect is stronger in hotter atmospheric models, for which the height of formation of both diagnostics coincides almost exactly as it is pushed to the chromosphere-TR boundary. The direct contribution to the H$\alpha$ line width from solely thermal Doppler broadening for the range of temperatures detected with ALMA (between 6,500 K to 12,000 K) is only about 0.025 \AA, an order of magnitude smaller than the observed variation of 0.3 \AA.

Indeed, the earlier interpretation of H$\alpha$
width as due essentially to thermal broadening \citep{2009A&A...503..577C} required a much larger range of electronic temperatures, all the way to T$_e \sim$ 60,000 K. With the reliable determination of T$_e$ now provided by ALMA, it appears necessary to revisit the original assumption of negligible changes due to radiative transfer effects \citep[essentially contained in the ``basal'' width of the lines in][]{2009A&A...503..577C}, at least for H$\alpha$. A strong correlation between H$\alpha$ width and \ion{Ca}{2} 8542 width and core intensity has also been observed in the dataset discussed in this paper,
and will be the subject of a future investigation. 
{We note that the smaller temperature range found here to explain the H$\alpha$ line width does not alter the need for a temperature-dependent microturbulent broadening to produce the observed distribution of widths of the \ion{Ca}{2} 8542 line.}

Of more general utility, we have demonstrated that the H$\alpha$ line width can be as useful and meaningful of an indicator of the temperature of the chromosphere and the initial rise into the transition region as those temperatures derived from ALMA millimeter intensities, at least in the range 6,500 -- 12,000 K. 
While it too suffers from the same changes in the heights of formation as the millimeter radiation, given the above caveats it can represent an easily accessible and straightforward diagnostic of chromospheric  temperatures in many regions of the chromosphere. The combination of H$\alpha$ line width and ALMA 3~mm observations has in effect allowed us to calibrate the line-widths in terms of brightness temperature (for observations of network regions and near disk center). {Future work should examine the nature of the relationship at different heliocentric angles or for different structures.} The values of the linear fit can be used to derive an approximate conversion of the line width to chromospheric 3 mm brightness temperature with an accuracy of better than 1000~K. Given the good matching between the 1-D models and the H$\alpha$ line width, the measured value could provide an efficient method to make better initial guesses of the input atmosphere for spectral inversions of chromospheric lines.

\acknowledgements
The National Solar Observatory (NSO) is operated by the Association of Universities for Research in Astronomy, Inc. (AURA), under cooperative agreement with the NSF. IBIS has been designed and constructed by the INAF/Osservatorio Astrofisico di Arcetri with contributions from the Universit\`a di Firenze, the Universit\`a di Roma “Tor Vergata”, {and upgraded with further contributions from NSO and Queens University Belfast.}
This paper makes use of the following ALMA data: ADS/NRAO.ALMA\#2016.1.01129.S. ALMA is a partnership of ESO (representing its member states), NSF (USA) and NINS (Japan), together with NRC (Canada), MOST and ASIAA (Taiwan), and KASI (Republic of Korea), in cooperation with the Republic of Chile. The Joint ALMA Observatory is operated by ESO, AUI/NRAO and NAOJ. 
{The National Radio Astronomy Observatory is a facility of the National Science Foundation operated under cooperative agreement by Associated Universities, Inc.}
This research has made use of NASA's Astrophysics Data System as well as Rob Rutten's SDO alignment IDL package. The authors would like to express their gratitude for the support staff at the DST, especially Doug Gilliam, the observing staff at ALMA, especially Juan Cortes, and the ALMA support scientist, Tim Bastian. MEM was supported by the George Ellery Hale Graduate Student Fellowship {from the University of Colorado, Boulder. The authors would like to thank the anonymous referee for the careful review of the manuscript and valuable suggestions.}  This research utilized the Python libraries matplotlib \citep{matplotlib} and the NumPy computational environment \citep{numpy}.

\bibliography{alma_paper.bib}{}

\end{document}